\documentclass[letterpaper, 10pt, conference]{ieeeconf}  
\pdfoutput=1
\usepackage{adjustbox}
\usepackage{resizegather}
\usepackage{graphicx}
\usepackage{graphics} 
\usepackage{epsfig} 
\usepackage[abs]{overpic}
\usepackage{algorithmic}
\usepackage{epstopdf}
\usepackage{caption}
\usepackage{subcaption}
\usepackage{setspace}
\IEEEoverridecommandlockouts
\usepackage[utf8]{inputenc}
\usepackage{tikz}
\usetikzlibrary{arrows}
\usepackage{authblk}

\usepackage[ruled,longend]{algorithm2e}

\title{Time Complexity Analysis of an Evolutionary Algorithm for approximating Nash Equilibriums}

\author{
    Aadesh Salecha\\
    Department of Computer Science\\
    University of Minnesota\\
    salec006@umn.edu
}

\date{}

\begin{document}

\maketitle

\begin{abstract}
The framework outlined in \cite{ours} provides an approximation algorithm for computing Nash equilibria of normal form games. Since, $NASH$ is a well-known $PPAD$-complete problem, this framework has potential applications to other $PPAD$ problems. The correctness of this framework has been empirically validated on 4 well-studied 2x2 games: Prisoner's Dilemma, Stag Hunt, Battle and Chicken. In this paper we provide the asymptotic time-complexities for these methods and in particular verify that for 2x2 games the worst-case complexity is linear in the number of actions an agent can choose from.

\end{abstract}

\section{Introduction}

Kuan and Salecha~\cite{ours} presented an evolutionary algorithm that amongst other things, can be applied to approximating Nash Equillibria of normal form games. This paper provides an time-complexity analysis of their methods and therefore presumes that the reader is well acquainted with the framework laid out in \cite{ours}.

The authors of \cite{ours} summarize the motivations and significance of studying the complex behaviours that arise from the incentive structures described by games such as Prisoner's Dilemma in sections I. of their paper and therefore we will refrain from reiterating these motivations. \\ 
The problem of computing the Nash equilibrium of games has far-reaching consequences in everything from conflict resolution between countries to structuring of auctions. Nash Equilibria are the cornerstone of modern economic theory. Researchers have been studying Nash Equilibria for decades and have studied various aspects of it, including the complexity of computing it. 

In \cite{ours}, the authors leveraged genetic algorithms to propose a framework that can be used to study the evolution of populations of agents which all compete with each-other in an environment where the incentive structure is defined by a normal form game. This framework can also be applied as an algorithm (hence forth referred to as the \texttt{evol-sim} algorithm) to compute the Nash equilibrium of the game that the agents compete in. The researchers in this paper were able to compute strategies equivalent to the mixed Nash Equilibrium strategies for well-known 2x2 normal form games using this algorithm. If their techniques can be proven to be efficient and correct, then this might be a step-towards a viable approximation algorithm for Nash Equilibrium computation. By providing a time-complexity analysis of the \texttt{evol-sim} algorithm we aim to address the questions about the efficiency of this framework.

\section{Problem Classification}

In 2009, Daskalakis et al. made a series of breakthroughs in discovering the theoretical bounds of computing Nash Equilibria (NE). The problem of computing a NE for games of three of more players was show to be in the class of problems called $PPAD$ \cite{daskalakis2009complexity}. It was later extended to 2 player NE by Chen and colleagues~\cite{chen2006settling}. The $PPAD$ (Polynomial Parity Arguments on Directed graphs) is a subset of $TFNP$ (Total Function NP) which itself is a subset of the $FNP$ (Function NP) class. The $FNP$ class is a function problem equivalent of the decision problem class $NP$. $TFNP$ is the subset in $FNP$ in which every problem is guaranteed to have a solution. Therefore, it is no surprise that computing Nash Equillibria falls in this subset, because in 1951 John Nash published his famous proof which proved that there always exists a mixed NE for any finite problem~\cite{nash1951non}.

The problem of computing a NE has not only been show to belong to $PPAD$ it is in fact $PPAD$ complete, which is to say that there exists a reduction between any other problem in $PPAD$ and this problem~\cite{daskalakis2009complexity}. This is of particular importance as it means that studying NE and its computation can provide information that can be generalized to problems that belong to $PPAD$ as a whole.

\section{Correctness}

Kuan et al. describe an approach that can be used as an approximation algorithm for the problem of computing a NE. The algorithm makes no guarantees concerning the quality of the solution or how much it differs from the optimal solution. But, the authors do provide empirical results that corroborate that the framework accurately computes strategies that are equivalent to mixed NE strategies for the following 2x2 normal form games: 

\begin{itemize}
    \item Battle
    \item Chicken 
    \item Stag Hunt
    \item Prisoner's Dilemma
\end{itemize}

The authors also a discuss hyper-parameter settings that lead to Pareto-optimal strategy computation for Prisoners Dilemma.

\section{Time-Complexity Analysis} 

The authors make their implementation of this framework that was used to generate the results for~\cite{ours} publicly available at \texttt{https://github.com/jinhongkuan/evol-sim}. In this section we briefly summarize the implementation, highlight some of the variables that are important to consider in this implementation, and finally delve into its time complexity analysis. 

\subsection{Variables of Concern}

In this subsection, we define a few variables that we make use of throughout the analysis. 

\begin{itemize}
    \item Population: This is the set of agents, each of which is represented by a Stochastic Moore Machine, that is being evolved in one instance of a run of this framework.
    \item Number of agents ($P$): The carnality of Population, ie. the number of agents in the population
    \item Number of generations ($G$): Empirically set constant that governs the number of iterations that the population is evolved for
    \item State Size ($S$): Governs complexity of strategy representable by an agent. It is the number of decision nodes in each SMM of an agent.
    \item Number of actions ($A$): Property of the problem/payoff matrix being solved. It is the number of distinct actions that an agent can choose to take, usually the number of columns/rows in a normal form game.
    \item Matrix Convergence Constant ($K$): An empirically set constant that governs the number of times the interaction matrix operations are performed. It is set so as to ensure the matrices reach near convergence values.
\end{itemize}

\begin{algorithm}
  $\textbf{Input:} \quad Population:$ Initial set of SMM agents,
  $\quad P:$ Number of agents in the $Population$\\
  $\textbf{Output:} \quad Population:$ Set of SMM agents that have been through $G$ generations of evolution \\
  
   \For{$k = 1 \dots G$}
   {
        $Population \gets Shuffle(Population)$ \\
        $fitness \gets Interaction(Population)$ \\
        $top\_parents \gets Reproductive\_Selection(Population, fitness)$ \\
        $Population \gets Survival\_Selection(Population, fitness)$ \\
        $Population \gets Population + Make\_children(top\_parents, mutator)$\\
   }
  \caption{\texttt{evol-sim} Algorithm}
  \label{algorithm1}
\end{algorithm}

\begin{algorithm}
  $\textbf{Input:} \quad Population:$ Set of agents, 
  $P:$ Number of agents in the $Population$,
  $K:$ Empirically found constant to allow agent interactions to reach horizon values

  $\textbf{Output:} \quad Scores:$ Fitness scores for each agent \\

   \For{$i = 1 \dots P$}
   {
        \For{$j = i+1 \dots P$}
        {
            \For{$k = 1 \dots K$}
            {
                Matrix operations to compute horizon values for scores of $Agent_i$ and $Agent_j$ based on the given payoff matrix
            }
        }
   }
   
  \caption{Interaction Function}
  \label{algorithm2}
\end{algorithm}

\subsection{Main Loop}

Algorithm \ref{algorithm1} summarizes the genetic algorithm driver. The $Population$ of agents are subjected to $G$ generations of evolution. Each generation consists of shuffling the population, followed by computing their fitness scores using the interaction function, followed up subjecting them to reproductive selection and survival selection. The ones chosen from reproductive selection are allowed to produce off springs that become a part of the population. There are a few other nuances here about extending the $Population$ to contain the offsprings and parents when the $Overlap$ variable is set to true that we ignore here as it does not affect the time-complexity analysis. 

\subsection{Interaction Function}

The most computationally expensive phase in the genetic algorithm driver is the computing of the fitness scores in the $Interaction$ function. This function, takes in the $Population$ of agents, and pits every agent against every other agent and has them go through iterated rounds of the game in question (for instance the paper discusses results related to Iterated Prisoner's Dilemma games). 

In order to efficiently simulate long iterations of these games between agents without actually simulating the games explicitly, this framework makes use of matrix computations that compute probability distributions that give accurate horizon values. We call the number of times that these matrix computations are performed as $K$. In the paper, the authors discuss this matrix computation in section B. of the appendix and also mention that they used $K$ = 5 for obtaining their results. The time-complexity of computing each of these $K$ matrix operations is $O(S \cdot A)$. 

Therefore, the total time-complexity of this function is $O(P^2 \cdot K \cdot S \cdot A)$ the $P^2$ comes from pitting of each agent against others and the $K \cdot S \cdot A$ comes from the matrix operations.

\subsection{Reproductive Selection}

The Reproductive Selection function is fairly straightforward. It implements one of two different paradigms of selection: 

\begin{itemize}
    \item Truncation: A constant number of top performing agents are selected to reproduce and make copies of themselves. Performance is governed by their fitness scores computed during the $Interaction$ function. Therefore, the time-complexity of this function is $O(P \cdot log P)$
    
    \item Roulette: Every agent has a probability of reproducing which is proportional to their fitness scores computed during the $Interaction$ function. Therefore the selection of a subset of the $Population$ to reproduce can be done in $O(P)$
\end{itemize}

Therefore, the time-complexity of performing reproductive selection is $O(P \cdot log P)$

\subsection{Survival Selection}

The Survival Selection function is similar to the $Reproductive\, Selection$ function. It implements one of two different paradigms of selection: 

\begin{itemize}
    \item Truncation: A constant number of top performing agents are selected to survive according to their fitness scores. Therefore, the time-complexity of this function is $O(P \cdot log P)$
    
    \item Uniform: Every agent has an equal probability of surviving. Therefore the selection of a subset of the $Population$ to reproduce can be done in $O(P)$
\end{itemize}

Therefore, the time-complexity of performing survival selection is $O(P \cdot log P)$

\subsection{Make\_Children}

This function takes in the agents that have been selected to reproduce (called $top\_parents$) and then makes copies of them whilst introducing variation during the reproduction process using the novel mutation algorithm described in \cite{ours}. 

The time-complexity of this function is $O(P)$

\subsection{Final Time-Complexity}

Having computed the time-complexities of each of the functions in this framework. We can now combine these results to derive the time-complexity of the \texttt{evol-sim} algorithm.

The time complexity is \\ $= O(G\,\cdot\,P^2\,\cdot\,K\,\cdot\,S\,\cdot\,A)$

\subsection{Time-Complexity for Experiments}

When evaluating this framework on the {\it Prisoner's Dilemma, Stag Hunt, Battle and Chicken} the authors used the following values for the constants: \\ 
$G = 1000,\, P = 10,\, K = 5,\, S = 2$ (empirically derived constants)\\ 
and $A = 2$ because for the 4 games evaluated were all 2x2 games where each agent had 2 action choices. \\

Therefore, the time-complexity for these 2x2 games was $10^6 \cdot A$, which is linear in terms of the input. The theoretical complexity for evaluating other 2x2 games would remain the same, as long as the empirically derived constants still work for the given 2x2 payoff matrix. 

If the empirical constants remain along the same order of magnitude, then this framework would make for a very efficient approximation algorithm that would scale linearly in the problem size. But we suspect that as the complexity of the games being evaluated go up, these hyper-parameters, especially $K$ and $S$ will not scale well. More experimental work is needed to get data about how this framework scales to larger problems. This along with the creation of theoretical guarantees for the quality of an approximation would make this algorithm a good choice for computing Nash Equilibria.

\bibliography{ESS.bib}
\end{document}